\documentclass[journal=jacsat,manuscript=article]{achemso}

\usepackage[version=3]{mhchem} % Formula subscripts using \ce{}
\usepackage{color}
\usepackage{ulem} %cross out
\usepackage{multirow} % multicolumn table
\usepackage{textgreek}
\author{Yossarian Liebsch}
\affiliation{Fakult\"at f\"ur Physik and CENIDE, Universit\"at Duisburg-Essen, 47057 Duisburg, Germany}
\email{yossarian.liebsch@uni-due.de}
\phone{+49 (203) 379 2321}

\author{Umair Javed}
\affiliation{Faculty of Physics, University Vienna, -1090 Vienna, Austria}

\author{Lucia Skopinski}
\affiliation{Fakult\"at f\"ur Physik and CENIDE, Universit\"at Duisburg-Essen, 47057 Duisburg, Germany}

\author{Leon Daniel}
\affiliation{Fakult\"at f\"ur Physik and CENIDE, Universit\"at Duisburg-Essen, 47057 Duisburg, Germany}

\author{Franziska Appel}
\affiliation{Fakult\"at f\"ur Physik and CENIDE, Universit\"at Duisburg-Essen, 47057 Duisburg, Germany}

\author{Radia Rahali}
\affiliation{CIMAP-GANIL, CEA-CNRS-ENSICAEN-UCN, Caen, 14076, France}

\author{Clara Grygiel}
\affiliation{CIMAP-GANIL, CEA-CNRS-ENSICAEN-UCN, Caen, 14076, France}

\author{Henning Lebius}
\affiliation{CIMAP-GANIL, CEA-CNRS-ENSICAEN-UCN, Caen, 14076, France}

\author{Carolin Frank}
\affiliation{Fakult\"at f\"ur Physik and CENIDE, Universit\"at Duisburg-Essen, 47057 Duisburg, Germany}

\author{Lars Breuer}
\affiliation{Fakult\"at f\"ur Physik and CENIDE, Universit\"at Duisburg-Essen, 47057 Duisburg, Germany}

\author{Leon Kirsch}
\affiliation{GSI Helmholtzzentrum für Schwerionenforschung, Planckstr. 1, 64291, Darmstadt, Germany}

\author{Frieder Koch}
\affiliation{GSI Helmholtzzentrum für Schwerionenforschung, Planckstr. 1, 64291, Darmstadt, Germany}

\author{Jani Kotakoski}
\affiliation{Faculty of Physics, University Vienna, -1090 Vienna, Austria}

\author{Marika~Schleberger}
\affiliation{Fakult\"at f\"ur Physik and CENIDE, Universit\"at Duisburg-Essen, 47057 Duisburg, Germany}

\title[Substrate-dependent pore formation in molybdenum disulfide monolayers under ion irradiation]
  {Substrate-dependent pore formation in molybdenum disulfide monolayers under ion irradiation}

\abbreviations{IR,NMR,UV}

\begin{document}

%%%%%%%%%%%%%%%%%%%%%%%%%%%%%%%%%%%%%%%%%%%%%%%%%%%%%%%%%%%%%%%%%%%%%
%% The "tocentry" environment can be used to create an entry for the
%% graphical table of contents. It is given here as some journals
%% require that it is printed as part of the abstract page. It will
%% be automatically moved as appropriate.
%%%%%%%%%%%%%%%%%%%%%%%%%%%%%%%%%%%%%%%%%%%%%%%%%%%%%%%%%%%%%%%%%%%%%
%\begin{tocentry}

%Some journals require a graphical entry for the Table of Contents.
%This should be laid out ``print ready'' so that the sizing of the
%text is correct.

%Inside the \texttt{tocentry} environment, the font used is Helvetica
%8\,pt, as required by \emph{Journal of the American Chemical
%Society}.

%The surrounding frame is 9\,cm by 3.5\,cm, which is the maximum
%permitted for  \emph{Journal of the American Chemical Society}
%graphical table of content entries. The box will not resize if the
%content is too big: instead it will overflow the edge of the box.

%This box and the associated title will always be printed on a
%separate page at the end of the document.

%\includegraphics[]{images/TOC.png}
%\label{For Table of Contents Only}

%\end{tocentry}

%%%%%%%%%%%%%%%%%%%%%%%%%%%%%%%%%%%%%%%%%%%%%%%%%%%%%%%%%%%%%%%%%%%%%
%% The abstract environment will automatically gobble the contents
%% if an abstract is not used by the target journal.
%%%%%%%%%%%%%%%%%%%%%%%%%%%%%%%%%%%%%%%%%%%%%%%%%%%%%%%%%%%%%%%%%%%%%
\begin{abstract}
Ion irradiation is a versatile tool for nanostructuring surfaces, yet the roles of energy deposition and dissipation at the surface and in ultrathin materials remain poorly understood. In this study, we investigate nanopore formation in monolayer MoS$_2$ on different substrates under irradiation of highly charged ions (HCIs) and swift heavy ions (SHIs): two types of ions that, despite having vastly different kinetic energies, interact primarily with the electronic system of the target. Using scanning transmission electron microscopy, we quantify pore radii and pore formation efficiencies for suspended MoS$_2$, MoS$_2$ on SiO$_2$, bilayer MoS$_2$ and MoS$_2$ on gold. Both pore size and pore formation efficiency exhibit a pronounced dependence on the type of substrate. Pores are largest and most frequent in MoS$_2$ on SiO$_2$, while the gold substrate massively quenches pore formation. The results indicate that the observed pore dimensions under both HCI and SHI irradiation are consistent with a central role of substrate and interface-dependent electronic dissipation pathways.

%Keywords: ion irradiation, molybdenum disulfide, substrate effects, pore formation, molecular dynamics simulations.

%
\end{abstract}

%%%%%%%%%%%%%%%%%%%%%%%%%%%%%%%%%%%%%%%%%%%%%%%%%%%%%%%%%%%%%%%%%%%%%
%% Start the main part of the manuscript here.
%%%%%%%%%%%%%%%%%%%%%%%%%%%%%%%%%%%%%%%%%%%%%%%%%%%%%%%%%%%%%%%%%%%%%
\section{Introduction}
Ion beams provide a controllable route to engineer defects in two-dimensional materials, enabling property tuning from doping to nanopore formation \cite{ElSaid.2012, Frost.2008, Vogel.2007, Akcoltekin.2007}. Because kinetic energy, charge state, and mass can be varied over wide ranges, ion irradiation offers a large parameter space for nanostructuring. Achieving predictive control requires a detailed understanding of ion–solid interaction and post-impact energy dissipation \cite{Telkhozhayeva.2024}. While bulk ion–solid interactions are well described across many energy regimes \cite{Rutherford.1911,Bohr.1913, LINDHARD.1963}, the interaction with surfaces and ultrathin targets—particularly for highly charged ions (HCIs) and swift heavy ions (SHIs)—remains less complete.

The emergence of two-dimensional (2D) materials has intensified interest in ion–surface interactions \cite{Novoselov.2005, Lehtinen.2010, Krasheninnikov.2010, Schleberger.2018, Liebsch.2026}. Owing to their atomic thickness, 2D materials combine outstanding mechanical properties with promising (opto-)electronic and catalytic functionality \cite{Kim.2019, Cao.2018, Cheng.2019, Madau.2018, Fruehwald.2025}. Ion irradiation has proven effective for defect engineering across regimes: low-energy ions enable implantation with near-atomic precision \cite{Junge.2022, Lin.2021, Auge.2022}, keV ions primarily create point defects in monolayers \cite{Pan.2014, GhorbaniAsl.2017}, and HCIs add potential energy that can drive larger defect complexes \cite{Kozubek.2019}. At MeV energies, damage is dominated by electronic excitation and can be tuned by the irradiation geometry (e.g., grazing incidence) \cite{Akcoltekin.2011, Madau.2017}.

While fundamental studies of ion irradiation-induced defect formation have largely focused on suspended 2D materials~\cite{Schwestka.2020, Creutzburg.2020, Kozubek.2019, Liebsch.2026}, practical applications typically require the 2D material to be supported by a substrate. Direct investigation of ion-irradiated supported 2D materials, however, remains experimentally challenging for several reasons. High-resolution STEM generally requires freestanding membranes, while atomic-resolution AFM and STM demand exceptionally clean surfaces and can further complicate interpretation, since both topographic and electronic contrast may contain contributions from the substrate as well as from the 2D layer~\cite{Hopster.2013, Zhang.2022, Villarreal.2024, Standop.2013}. As a result, many studies of supported systems rely on Raman, photoluminescence, or electrical measurements, which are sensitive to electronic and strain-related changes but do not directly resolve the atomic structure~\cite{Ochedowski.2015, Sleziona.2024}.

This leaves a gap in our understanding of how ion-induced defect formation differs between suspended and substrate-supported 2D materials. We address this gap by directly measuring substrate effects on defect formation in MoS$_2$
under HCI and SHI irradiation. HCIs and SHIs represent two distinct irradiation regimes with vastly different kinetic energies and energy-deposition profiles. HCIs release their potential energy predominantly through ultrafast charge exchange and Auger-type processes near the surface~\cite{Winter.2001,Wilhelm.2017}, whereas SHIs deposit energy continuously along their trajectory via electronic stopping 
$S_e$. Despite these differences, both ion types couple primarily to the electronic system of the target, such that subsequent energy transfer to the lattice is governed by electron–phonon coupling~\cite{Wang.2014,Dufour.2017}. This makes HCI and SHI irradiation a useful pair of probes for testing how substrate-controlled electronic dissipation influences defect formation in 2D materials.

Here we directly quantify how substrate coupling influences electronically driven defect formation in monolayer MoS$_2$. Using scanning transmission electron microscopy, we measure pore radii and pore formation efficiencies after irradiation in several configurations (suspended, supported on SiO$_2$, bilayer, and metal-supported samples). We compare HCIs and SHIs as complementary excitation regimes with distinct energy-deposition profiles but a shared dependence on post-impact electronic energy dissipation. Owing to this common feature, we expect the substrate may influence pore formation in a qualitatively similar way for both ion types. %This approach provides atomic-scale defect statistics that link ion energy deposition to substrate-dependent dissipation pathways governing pore formation. 

\section{Results and discussion}
\subsection{Highly Charged Ions}
\begin{figure}[htbp]
\centering
\includegraphics[width=\textwidth,keepaspectratio]{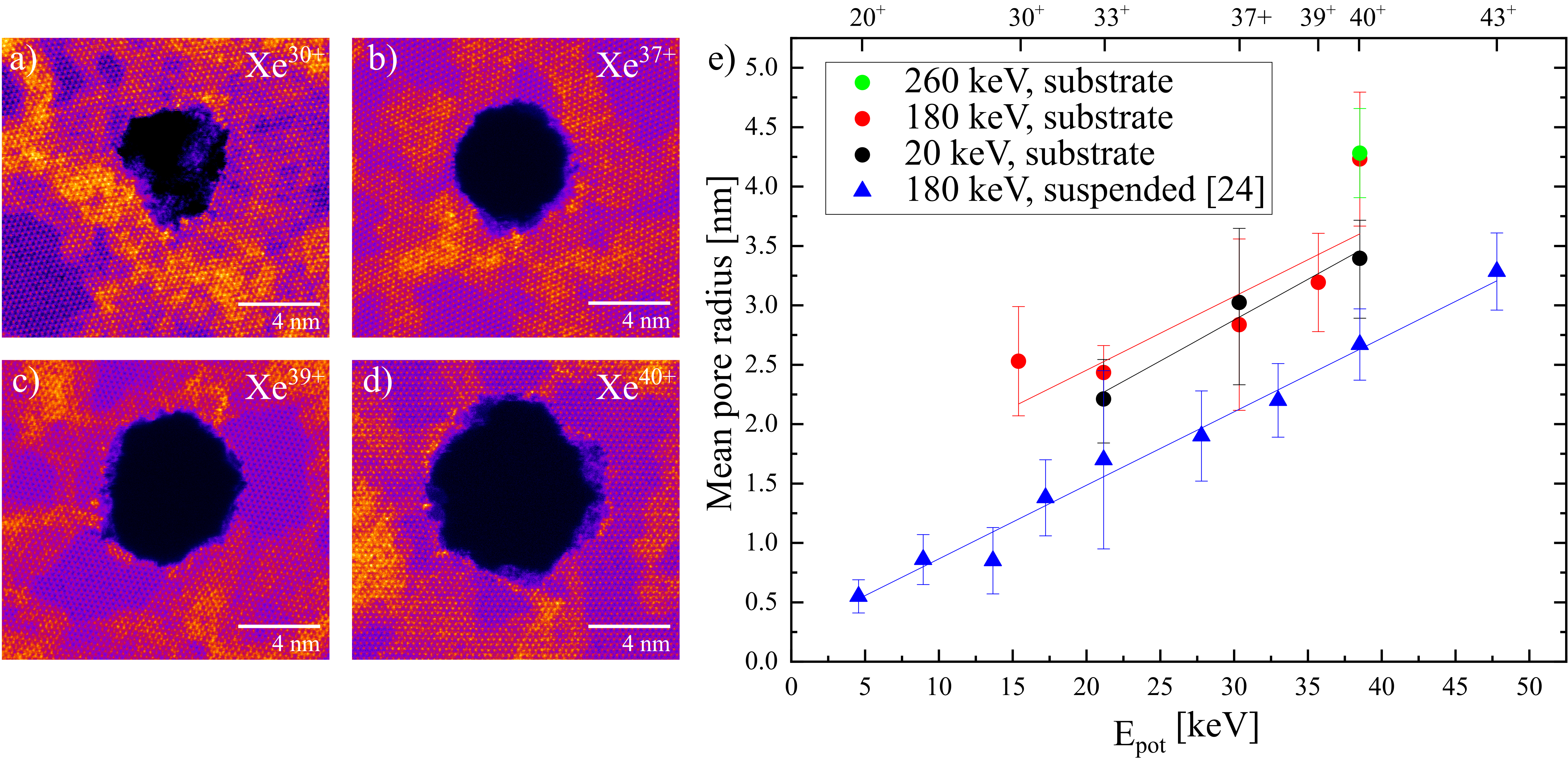}
\caption{(a-d) False-color STEM images of pores in single-layer MoS$_2$ created by Xe ions with different charge states at 180~keV. Orange structures are caused by hydrocarbon contamination. (e) Mean pore radii of SL-MoS$_2$ on SiO$_2$/Si substrate irradiated with highly charged Xe ions at different kinetic energies (green, red, black). Pore radii are larger when MoS$_2$ is irradiated on SiO$_2$ compared to the suspended configuration (blue, data by Kozubek \textit{et al.}) \cite{Kozubek.2019}.}
\label{fig3}
\end{figure}

In order to assess substrate effects under HCI irradiation, we compare monolayer MoS$_2$
on SiO$_2$/Si with suspended MoS$_2$ reported by Kozubek et al. \cite{Kozubek.2019}. Using the same ion species and kinetic energies as in that work, we vary the charge state to deliver between $15-40$\,keV of potential energy to the surface. In contrast to earlier AFM studies on supported samples that suggested pronounced topographic changes but could not resolve the defect structure \cite{ElSaid.2008,Hopster.2014}, our STEM analysis shows that irradiation on SiO$_2$ produces well-defined, pore-shaped defects with no obvious long-range lattice distortion of the surrounding material (see Fig.S1), consistent with the pore morphology previously reported for suspended MoS$_2$\cite{Kozubek.2019}.
Representative pores are shown in Fig.\,\ref{fig3}a-d.

We quantified pore-size distributions by measuring ($\sim$\,150) pores. Mean pore radii are summarized in Fig.\,\ref{fig3}e. For both supported and suspended samples, the mean radius increases approximately linearly with the projectile potential energy. Within uncertainty, the slopes are similar for supported and suspended ($0.06-0.07$ nm/keV), indicating comparable scaling with charge state. Pores on SiO$_2$ are systematically larger by $\sim1$\,nm across all charge states. In addition, within our experimental resolution we observe no significant dependence of pore radius on kinetic energy over the investigated range.

This observation is consistent with sputtering experiments of SL-MoS$_2$ carried out by Skopinski \textit{et al.}, who found that the sputtering yield of Mo is significantly more sensitive to changes in the charge state of the HCI rather than its kinetic energy~\cite{Skopinski.2023}. These observations suggest that nuclear sputtering contributes only weakly under the present conditions and pore formation is dominated by the deposition and dissipation of the HCI's potential energy.

Charge-exchange studies for HCIs on 2D materials carried out by Creutzburg \textit{et al.} and Niggas \textit{et al.} found that the majority of the HCIs potential energy ($\sim 80-90$\,\%) is deposited \textit{via} Auger processes in the electronic system of the first layer of the target 2D material, independent of the materials electronic properties~\cite{Creutzburg.2020,Niggas.2022}. Linking the charge exchange to pore formation, Grossek \textit{et al.} introduced a model for nanopore formation in graphene, that centers around the target's charge carrier mobility~\cite{Grossek.2022}. They argue that the high electron mobility of graphene leads to a fast charge dissipation in the 2D material, effectively suppressing pore formation. As a consequence, HCI irradiation does not create pores in graphene (typically $\geq\mu=60.000$~cm$^2$~V$^{-1}$s$^{-1}$ for suspended graphene \cite{Bolotin.2008}), but very likely in MoS$_2$, which has a significantly smaller electron mobility (typically $\mu\approx1-10$~cm$^2$~V$^{-1}$s$^{-1}$~for suspended MoS$_2$~\cite{Jariwala.2013,Kim.2016}).

Building on the model introduced by Grossek et al., in which charge carrier mobility governs the efficiency of pore formation under HCI irradiation \cite{Grossek.2022}, the larger pores observed in MoS$_2$ on SiO$_2$ can be interpreted in terms of substrate-modified electronic energy dissipation within the MoS$_2$ layer. For an insulating substrate, out-of-plane dissipation remains limited. However, substrate-induced disorder and charge trapping are expected to increase carrier scattering, thereby limiting lateral spreading of the non-equilibrium electronic excitation. This is known from several studies on charge transport in MoS$_2$ that report higher carrier mobility in suspended monolayers than in MoS$_2$ on SiO$_2$ \cite{Chen.2018, Yan.2026}. Reduced mobility can slow lateral charge redistribution, which may increase the spatial localization of the transient excitation and thus raise the local electronic energy density. This in turn may increase the local energy density available for lattice excitation. We note that this interpretation is inferential---the dissipation pathway is not measured directly---but it is consistent with the pore size enhancement observed under both HCI and SHI irradiation (see Fig.\,\ref{fig:Daten} a), suggesting the effect is not specific to one excitation regime.

\begin{figure}[htbp]
\centering
\includegraphics[width=\textwidth,keepaspectratio]{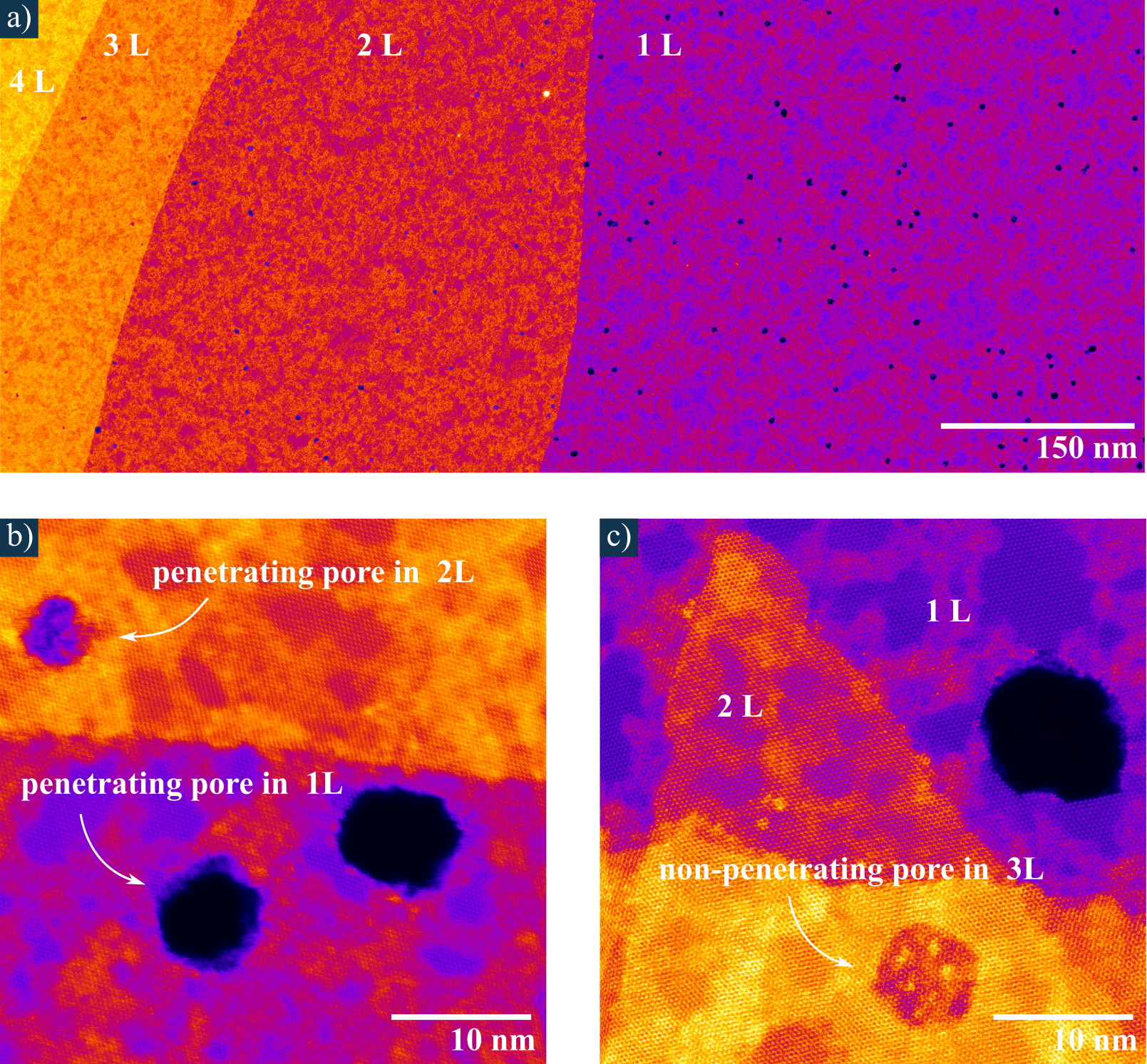}
\caption{(a) STEM images of MoS$_2$ on SiO$_2$ with different layer numbers, irradiated with 180\,keV Xe$^{37+}$ ions. (b) Fewer and smaller pores are observed in bilayer MoS$_2$ (2L) compared to monolayer on SiO$_2$ (1L). (c) In trilayer MoS$_2$, pores are predominately non-penetrating. }
\label{fig:multilayer}
\end{figure}

The preceding discussion focused on lateral (in-plane) dissipation pathways, treating out-of-plane dissipation as negligible for an insulating substrate. If, however, a semiconducting or metallic substrate is used, this additional dissipation channel could strongly influence pore formation. Multilayer MoS$_2$ provides a controlled way to introduce out-of-plane dissipation without changing the substrate material. Although adjacent layers are coupled only weakly via van-der-Waals forces, a second (or third) MoS$_2$ layer can act as an additional reservoir for charge and excitation redistribution relative to an isolated monolayer, offering a simplified analogue of out-of-plane dissipation into a semiconducting environment.

To test the role of out-of-plane electronic energy dissipation, we exploit regions of different thickness in CVD-grown MoS$_2$ (Fig.\,\ref{fig:multilayer}a). Compared to monolayer regions (1L), bilayer MoS$_2$ (2L) shows both reduced pore radii and reduced pore formation efficiency (Fig.\,\ref{fig:multilayer}b). In trilayer regions (3L), fully penetrating pores are strongly suppressed and partially penetrating (non-through) pores occur (Fig.\,\ref{fig:multilayer}c). These observations indicate that out-of-plane electronic energy dissipation is present but limited in magnitude: coupling to a second layer is sufficient to modify pore formation, yet the available excitation appears insufficient to efficiently drive a fully penetrating defect through three layers.

The observed anisotropy of the defects (radius\,$\approx2.2$\,nm; depth\,$\approx1.5$\,nm) likely reflects the strongly different in-plane and out-of-plane dissipation pathways in layered MoS$_2$~\cite{Ding.2015,Lee.2021}, which disfavor spherical damage volumes.

\subsection{Swift Heavy Ions}
\begin{figure}[htbp]
\centering
\includegraphics[width=\textwidth,keepaspectratio]{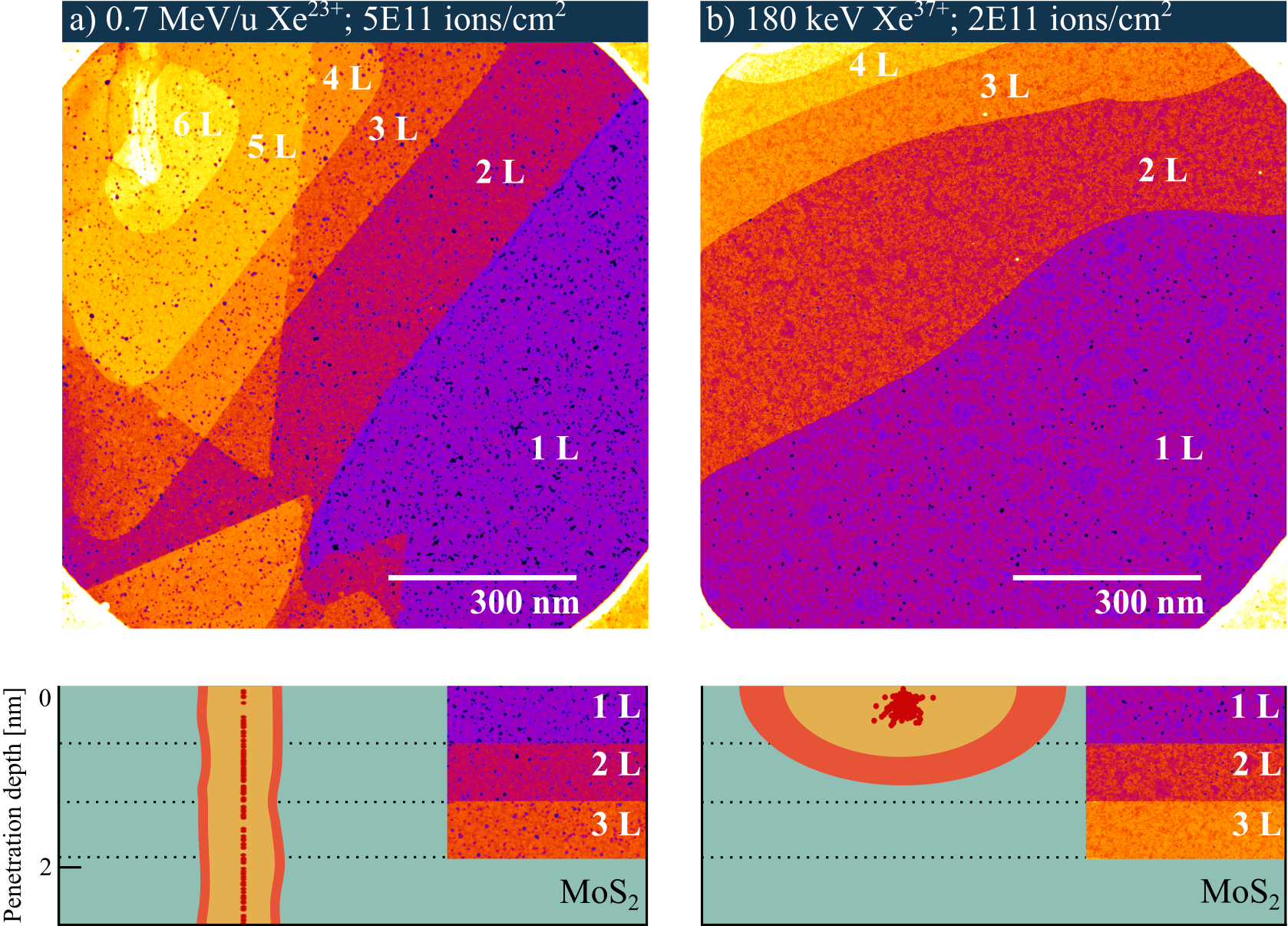}
\caption{Comparison of SHI-irradiated (a) and HCI-irradiated (b) MoS$_2$ reveals similar pore dimensions in 1L, but, as expected, vastly different penetration behavior. Schematics below each image illustrate energy deposition profiles of the different irradiation types. Note that these illustrations are visual representations and not to scale.
}
\label{fig:HCI_vs_SHI}
\end{figure}

In addition to HCI irradiation, we investigate samples irradiated with SHIs. SHIs deposit energy continuously along their trajectory via electronic stopping 
$S_e$, producing excitation and ionization of the target’s electronic system (Fig.\,\ref{fig:HCI_vs_SHI}) \cite{Toulemonde.1993,Ziegler.2010}. This extended deposition profile provides a complementary probe of substrate-controlled electronic energy dissipation. Based on previous results on suspended MoS$_2$, the SHIs used in this study are expected to deposit in the order of $\sim10$\,keV (0.7 MeV/u Xe$^{23+}$) and $\sim20$\,keV (4.8 MeV/u Au$^{25+}$) in a monolayer~\cite{Liebsch.2026}.

\begin{figure}[htbp]
\centering
\includegraphics[width=\textwidth,keepaspectratio]{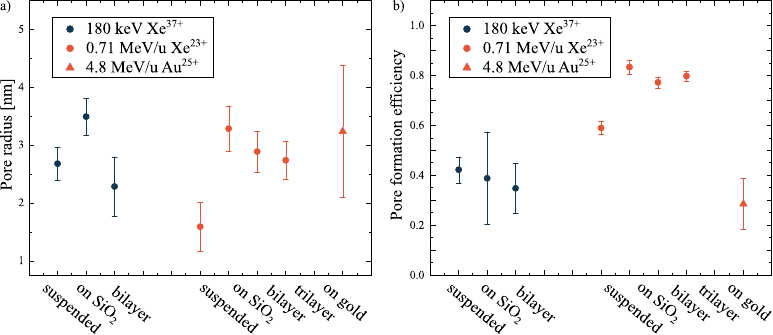}
\caption{Mean pore radii (a) and pore formation efficiencies (b) of HCI and SHI-irradiated MoS$_2$ in different configurations. Error bars of mean pore radii are the standard deviation, while uncertainty in efficiency arises from pore counting and fluence determination in all cases, and for HCI-irradiated supported samples additionally from partial-area irradiation, transfer-related misalignment (see Fig.S3). }
\label{fig:Daten}
\end{figure}

In Fig.\,\ref{fig:HCI_vs_SHI}, representative multilayer regions after HCI and SHI irradiation are shown. Pores in monolayers are of comparable size, while the penetration behavior differs strongly. Quantitative pore radii and pore formation efficiencies for all configurations are summarized in Fig.\,\ref{fig:Daten}. For both ion types, monolayer MoS$_2$ on SiO$_2$ exhibits the largest pores. However, thickness affects the two regimes differently: under SHI irradiation pore size and efficiency depend only weakly on layer number, whereas HCI-induced pore formation attenuates rapidly with thickness. This contrast is consistent with SHIs transferring electronic energy along an extended track deep into the target, while HCIs release their potential energy predominantly near the surface during a single neutralization event.  

The pronounced difference in size and efficiency between suspended and supported MoS$_2$ under SHI irradiation is consistent with the strong impact-parameter dependence of electronic energy deposition in an atomically thin target \cite{Liebsch.2026}. In suspended monolayers, a significant fraction of ion passages yields local energy densities that remain below the threshold for pore formation, depending on the trajectory relative to atomic sites. Adding a substrate and/or additional layers reduces the likelihood of such “sub-threshold” events because energy deposition and subsequent dissipation are distributed across coupled layers. Moreover, we previously showed that pore formation under SHI irradiation is strongly quenched by energy loss through particles escaping from both sides of a suspended membrane \cite{Liebsch.2026}. A substrate hinders this loss channel by restricting escape to one side, further increasing the effective energy available for damage formation.

For HCIs, the initial interaction is governed by near-surface charge exchange within a nm-scale zone spanning multiple atoms~\cite{Yamamura.1995,Insepov.2006}. This may reduce sensitivity to atomic-scale impact parameter compared to the SHI impact.

We observe that the pore formation efficiencies differ for the two ion types. Pore formation efficiency is here defined as the number of pores per incident ion. It is measured using large STEM images containing $\sim 100$ pores (see Supporting Information). HCIs yield generally lower efficiencies ($\eta\sim0.4$) for the same target configuration than SHIs ($\eta\sim0.6-0.9$). HCI efficiencies carry larger systematic uncertainty, arising from a fluence gradient across the beam spot and the transfer-related misalignment (Fig. S3), and should therefore be interpreted with caution. An additional source of local suppression is hydrocarbon contamination: ambient exposure produces hydrocarbon coverage on 2D materials \cite{Palinkas.2022}, and we observe indications of locally reduced pore formation in contaminated regions (Fig. S5), which likely lowers the apparent efficiency in the case of HCI.

For SHI irradiation, efficiency data are more meaningful. Efficiency in supported MoS$_2$ approaches saturation ($\eta\sim0.8$), whereas efficiency in the suspended monolayer remains well below saturation $\eta\approx0.6$. This behavior can, again, be traced back to sub-threshold energy transfer impacts that can occur in monolayers~\cite{Liebsch.2026}. The notably low efficiency observed for MoS$_2$ on Au ($\eta\approx$ 0.28) stands apart from this trend and will be discussed further below.

A remaining question is whether pore formation can be mitigated by the right choice of substrate. Schwestka \textit{et al.} demonstrated that graphene/MoS$_2$ heterostructures can be resistant to HCI-induced pore formation when MoS$_2$ is covered by graphene, as charge exchange and potential energy deposition primarily occurs within the graphene \cite{Schwestka.2020}. Importantly, when the heterostructure is inverted such that the HCI reaches MoS$_2$ first, pores are still created in MoS$_2$. 

This behavior is consistent with the weak coupling of different layers in van-der-Waals heterostructures. Although graphene is an excellent in-plane conductor, the out-of-plane electrical (and thermal) coupling across the graphene/MoS$_2$ interface is comparatively weak~\cite{Ding.2016,Liao.2018}. Simulations by Schwestka et al. suggest that the out-of-plane conductance across the graphene/MoS$_2$ interface is two orders of magnitude smaller than the in-plane conductance of the individual layers \cite{Schwestka.2020}. Consistent with limited charge transfer between MoS$_2$ and graphene, no strong PL quenching through loss of photoexcited charge carriers from the MoS$_2$ into the graphene is observed in graphene/MoS$_2$ heterostructures~\cite{Huo.2015}.

In contrast, coupling between 2D semiconductors and Au can be substantially stronger, which underlies the success of Au-assisted exfoliation and the formation of large-area, clean interfaces ~\cite{Desai.2016,Pollmann.2021}. Motivated by this, we irradiated MoS$_2$ exfoliated on Au to test whether a metallic substrate with strong interfacial coupling can act as an efficient sink for transient electronic excitation and charge, thereby suppressing pore formation under electronically dominated ion impact. 

In contrast to the samples discussed above, for which MoS$_2$ was transferred after irradiation using only water, transfer from Au substrates to TEM grids requires chemical removal of the gold support using an iodine/iodide etchant (KI/I$_2$). Consequently, unlike in the previous case, the ion-induced pores cannot be assumed to remain unaffected during transfer~\cite{Bala.2022}, as they were in the previous case. For this reason, the pore formation efficiency is expected to be the more reliable observable for a quantitative discussion of this substrate, whereas pore dimensions can only be discussed qualitatively. 

\begin{figure}[htbp]
\centering
\includegraphics[width=\textwidth,keepaspectratio]{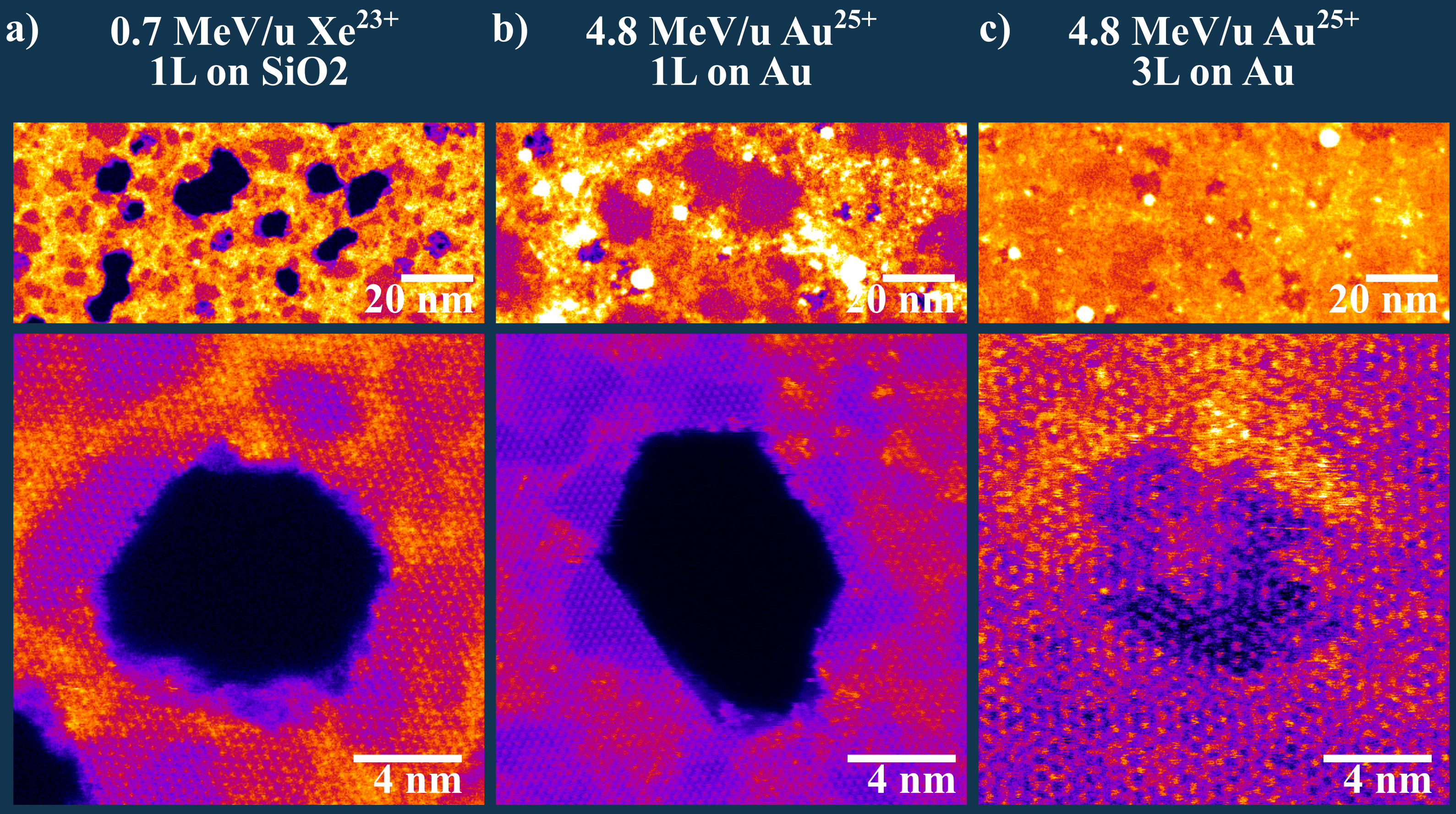}
\caption{Comparison of SL-MoS$_2$ irradiated on SiO$_2$ (a) and on gold substrate (b). On gold, pore formation efficiency is strongly reduced compared to MoS$_2$ on SiO$_2$. In addition, the expected increase in pore dimensions due to the difference in predicted stopping power between the two irradiations is not observed. (c) In trilayer MoS$_2$ on Au, only a single layer is penetrated, as evidenced by the Moiré pattern (caused by bilayer MoS$_2$) present on top of the pore. Bright, white spots in the STEM images are caused by dense agglomerates that can either stem from irradiation (Mo) or the transfer (Au). The lighter orange color of the trilayer image compared to monolayer originates from thickness-dependent scattering of electrons. } 
\label{fig:MoS2Au}
\end{figure}
Fig.\,\ref{fig:MoS2Au} compares SHI-irradiated MoS$_2$ on SiO$_2$ (a) with MoS$_2$ irradiated on Au (b). The two irradiations use different ions: the electronic stopping power of 4.8 MeV/u Au$^{25+}$ in suspended MoS$_2$ is approximately double that of 0.71 MeV/u$^{23+}$ Xe~\cite{Liebsch.2026}, such that larger pores would be expected on Au in the absence of substrate effects, as previously shown for suspended MoS$_2$ \cite{Liebsch.2026}. Instead, pores of similar dimension to those on SiO$_2$ are observed (Fig.\,\ref{fig:Daten} a). The pores on Au additionally exhibit a broad size distribution and an unusual morphology (see Fig.S6 and Fig.S7), with many showing straight, lattice-aligned edges consistent with faceted etch pits \cite{ElSaid.2010}. We interpret this as evidence of the aforementioned influence of the etchant during transfer, which may bias the measured pore sizes toward larger values~\cite{Bhowmik.2025}.

All irradiations shown in Fig.\,\ref{fig:MoS2Au} were performed at the same fluence ($5\times10^{11}$\,cm$^{-2}$). The pore formation efficiency of MoS$_2$ on Au is strongly reduced ($\eta\approx0.28$) compared with MoS$_2$ on SiO$_2$ ($\eta\approx0.83$). In trilayer regions of the same sample (Fig.\,\ref{fig:MoS2Au} c), a new type of defect is observed: here, only one of the three layers is perforated. This is apparent from the Moirè pattern observed on top of defects (c) that only appears if at least two layers of MoS$_2$ are present. These non-penetrating defects in trilayer MoS$_2$ occur at a much higher rate ($\eta\approx0.89$) than penetrating-pores in the monolayer regions. We interpret this contrast as evidence that the Au substrate acts as an efficient sink for electronic excitation generated in the first two MoS$_2$ layers near the interface, reducing the electronic energy density retained in MoS$_2$ below the threshold for pore formation. The high efficiency in the trilayer case is consistent with this picture: in the outermost layer, sufficiently far from the Au interface, dissipation into the substrate is reduced and the local excitation can exceed the formation threshold. As discussed for SHI-irradiated suspended MoS$_2$~\cite{Liebsch.2026}, pore formation efficiency is highly sensitive to threshold shifts, explaining why the efficiency changes drastically between 1L and 3L while pore sizes remain comparable.

The inferred efficient out-of-plane dissipation from MoS$_2$ into Au is consistent with the well-known strong MoS$_2$–Au interfacial coupling. Our own Raman and PL characterization of the MoS$_2$/Au samples (see Fig.S2) shows splitting of the out-of-plane $A_{1g}$ mode and strong PL quenching relative to MoS$_2$ on SiO$_2$, indicating that the interface modifies both lattice dynamics and carrier recombination in the as-prepared samples. The microscopic origin of this coupling has been attributed to Fermi-level pinning arising from interface-dipole formation and the appearance of Mo d-orbital gap states due to metal–S interactions at the interface~\cite{Gong.2014, Pollmann.2021}. Whether these equilibrium coupling mechanisms directly govern the efficiency of transient charge transfer under ion impact remains an open question, but the consistency between the strong equilibrium coupling and the observed suppression of pore formation supports the interpretation that Au provides an efficient dissipation channel for electronically driven damage.

The trends in pore size and pore formation efficiency observed here are robust, but the underlying dissipation pathways are inferred indirectly from defect statistics rather than measured directly. These limitations do not alter the main comparative conclusions, but they constrain the level of mechanistic interpretation that can be drawn from the present dataset.

\section{Conclusion}
We have shown that substrate coupling plays a decisive role in electronically driven pore formation in MoS$_2$ under both HCI and SHI irradiation. An insulating substrate increases pore size in comparison to suspended membranes, out-of-plane energy dissipation redistributes energy and suppresses pore formation in MoS$_2$ multilayers, and a strongly coupled metallic substrate (Au) markedly reduces the pore formation efficiency. Together, these trends indicate that, for ions that primarily interact with the electronic system of the target, pore evolution is governed less by the energy deposition mechanism than by how efficiently that electronic excitation is redistributed and converted into lattice motion under the constraints imposed by the interface. The presented substrate- and thickness-resolved pore statistics provide quantitative benchmarks for future modeling. The most promising next step is likely not a full atomistic treatment of the entire ion impact, but instead a targeted description of electronic energy dissipation across the interface between the 2D material and the substrate.

\section{Methods and Materials}
\subsection{Sample Preparation}
\begin{figure}[htbp]
\centering
\includegraphics[width=\textwidth,keepaspectratio]{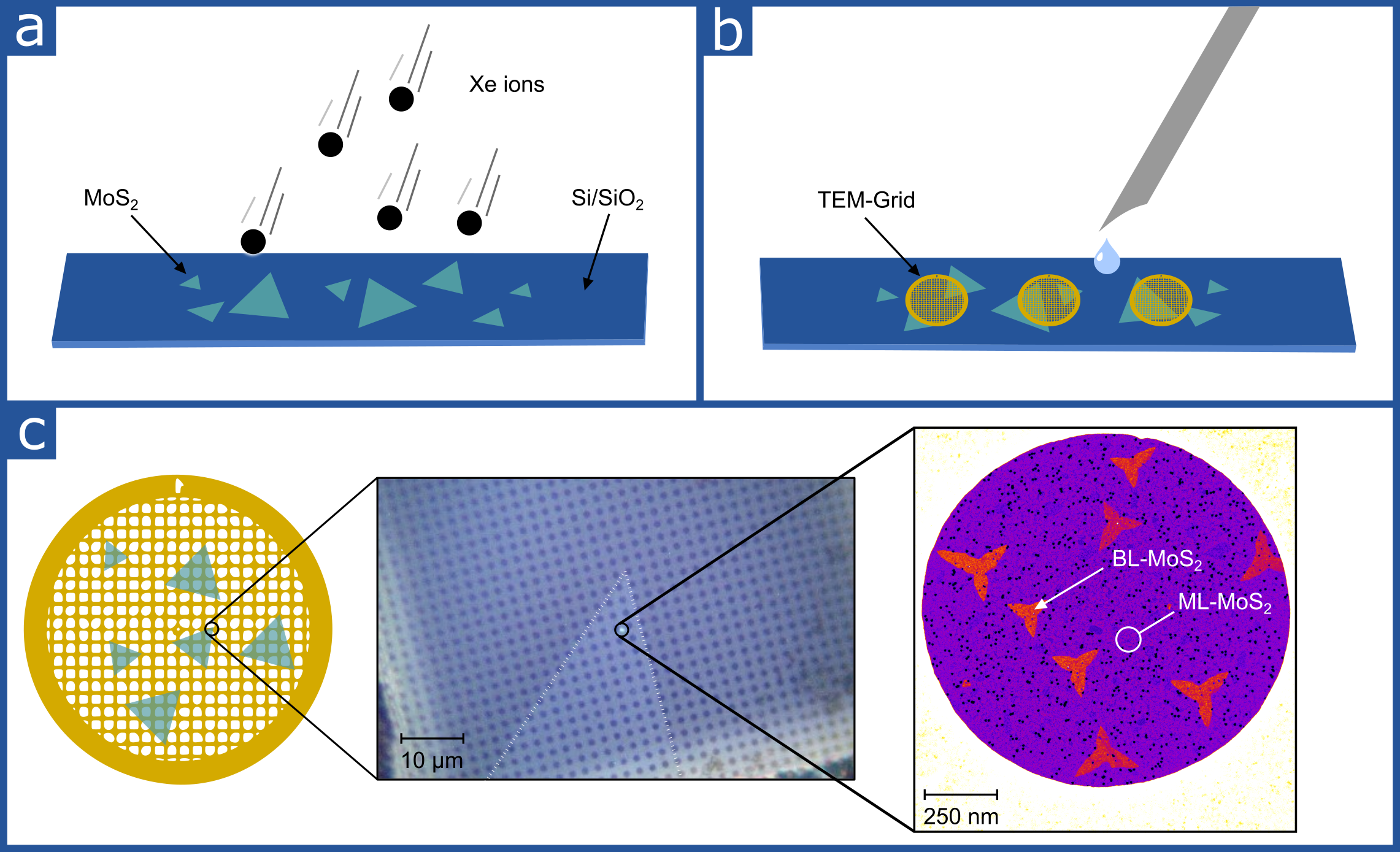}
\caption{Schematic of the irradiation ans subsequent STEM analysis of MoS$_2$ on SiO$_2$. First, CVD-grown MoS$_2$ on SiO$_2$/Si is irradiated with highly charged ions (a). After that irradiated MoS$_2$ is transferred polymer-free to a TEM-Grid (b). (c) Transferred flakes on the TEM-grid (left) are visible in an optical microscope (center). On the right, a STEM image of a single TEM-Grid pore, covered by monolayer (ML) and occasionally bilayer (BL) MoS$_2$ is shown.}
\label{fig1}
\end{figure}

A custom chemical vapor deposition (CVD) process is used to grow monolayer MoS$_2$. Substrates are prepared by placing a micro droplet of saturated ammonium heptamolybdate (AHM) onto 300~nm SiO$_2$/Si substrate. The substrates are then spin-coated with a 1\,\% sodium cholate solution that acts as a growth promoter. Substrates are then placed into the second zone of a three-zone furnace. In the first zone $\sim$ 60\,mg of sulfur is placed. An Argon flow of 500~sccm is used to purge the quartz tube containing the reagents and as means to transport the precursors. During the 30~min long process, the first zone reaches 150~$^{\circ}$C, while the second zone reaches 750~$^{\circ}$C. After the process has finished, the furnace is opened to allow for rapid cooling. The freshly grown MoS$_2$ flakes are irradiated on the SiO$_2$/Si substrate and subsequently transferred to a QUANTIFOIL© holey carbon grid with 1.2 µm holes. The transfer is performed by placing the TEM-grid onto the sample with the mesh facing the MoS$_2$ side. By placing a droplet of water near the grid, the water can intercalate between the flakes and the substrate, lifting off the MoS$_2$ in the process and pressing it against the QUANTIFOIL© holey carbon mesh. Sample preparation and characterization are schematically shown in Fig.\,\ref{fig1}. 

To investigate the irradiation of MoS$_2$ on Au, a substrate suitable for mechanical exfoliation is first prepared. A 5\,nm Ti adhesion layer is deposited onto SiO$_2$, followed by the deposition of a 25\,nm Au film. Immediately after deposition, MoS$_2$ is mechanically exfoliated onto the freshly prepared Au surface.

After irradiation on Au, the monolayers had to be transferred onto TEM grids for STEM characterization. For this purpose, a thin polystyrene (PS) layer is spin-coated onto the sample. The sample is then immersed in a preheated KI/I$_2$ etching solution (4\,g KI, 1.1\,g I, 150\,ml H$_2$O) at 40\,$^\circ$C. After approximately 1\,h, the Au layer has dissolved, allowing the PS film with the attached monolayers to be retrieved and cleaned in ultrapure water.

The film is then transferred onto a TEM grid and heated for 30\,min at 80\,$^\circ$C, followed by 1\,h at 130\,$^\circ$C. During this step, slow heating and cooling are essential to avoid damaging the TEM grid. Finally, the PS film is dissolved in toluene for 2\,h, with the solvent renewed once during the process. After retrieval, the grid is immersed in analytical-grade isopropanol as a final cleaning step and then left to dry.  
\subsection{Ion irradiation}  
Irradiation with highly charged Xenon ions was done at the HICS beamline at the University of Duisburg-Essen \cite{Peters.2009,Skopinski.2021,Skopinski.2023}. The ions with charge states $q$=28+ up to $q$=44+ are provided by an electron beam ion source (EBIS). Samples were irradiated under perpendicular incidence with a fluence of $\Phi$=\,5$\times$10$^{11}$\,cm$^{-2}$. At this fluence the density of pores is high enough to image them efficiently in the STEM, while it is low enough to avoid frequent occurrence of overlapping pores (see Supplementary Information). Irradiation was performed at three different kinetic energies (20\,keV, 180\,keV, 260\,keV) in order to evaluate the influence of the kinetic energy on the pore creation. As there was no evidence for differences between the two high energies, no additional irradiations at 260\,keV after the initial one were done. Irradiation with SHIs was performed at IRRSUD at GANIL, Caen (0.7 MeV/u $^{129}$Xe$^{23+}$) and at M-Branch at GSI, Darmstadt (4.8 MeV/u $^{197}$Au$^{25+}$). For both irradiations flux was kept below 1$\times$10$^{9}$\,s$^{-1}$\,cm$^{-2}$ to avoid thermal damage. Total fluence was again set to $\Phi$=\,5$\times$10$^{11}$\,cm$^{-2}$. All irradiations were under perpendicular incidence. 

\subsection{Imaging}
Scanning transmission electron microscopy (STEM) measurements were carried out in Vienna with an aberration corrected Nion UltraSTEM 100. Prior to imaging, the samples were heated at 170°C for around 10 h in vacuum to minimize the amount of surface contamination, and inserted into the microscope without air exposure \cite{mangler2022materials}. The images were recorded with medium angle annular dark field (MAADF) detector with a collection semiangle of 60-200 mrad. Acceleration voltage of electrons was 60 kV and a dwell time of 8 µs/px and a flyback time of 120 µs were used to record images with 1024$\times$1024 px. Images were recorded at fields of view of 32 nm for the pore sizes analysis and 512 nm for obtaining an overview. At least 150 pores were analyzed for each set of ion irradiation parameters. As most of the pores display round shapes, the area was converted to pore radii $r$ according to A = $\pi r{^2}$.

\subsection{Raman and PL spectroscopy}
Raman and PL spectroscopy were performed using a WITec alpha300 RA confocal Raman spectrometer. In all instances a green laser ($\lambda=$532\,nm) with a maximum power of 1\,mW was used. For Raman spectra a 1800 l\,mm$^{-1}$ grating was used, while a 300 l\,mm$^{-1}$ grating was used for PL recording.

%%%%%%%%%%%%%%%%%%%%%%%%%%%%%%%%%%%%%%%%%%%%%%%%%%%%%%%%%%%%%%%%%%%%%
%% The "Acknowledgement" section can be given in all manuscript
%% classes.  This should be given within the "acknowledgement"
%% environment, which will make the correct section or running title.
%%%%%%%%%%%%%%%%%%%%%%%%%%%%%%%%%%%%%%%%%%%%%%%%%%%%%%%%%%%%%%%%%%%%%
\begin{acknowledgement}

UJ and JK acknowledge funding from the Austrian Science Fund (FWF) within the Cluster of Excellence MECS (DOI: 10.55776/COE5). 
Y.L, M.S, L.S, L.D, C.F and L.B thank the Deutsche Forschungsgemeinschaft (DFG, German Research Foundation) for the support of this work under project numbers 272132938, 429784087 and 501495566, and through IRTG 2803 “2D Mature” (project number 461605777; P05), as well as project numbers 501495566 and 272132938.
Results presented here are partly based on experiments performed at the beam line M3 at the GSI Helmholtz\-Zentrum für Schwerionenforschung, Darmstadt (Germany) in the context of FAIR Phase-0.
\end{acknowledgement}

%%%%%%%%%%%%%%%%%%%%%%%%%%%%%%%%%%%%%%%%%%%%%%%%%%%%%%%%%%%%%%%%%%%%%
%% The appropriate \bibliography command should be placed here.
%% Notice that the class file automatically sets \bibliographystyle
%% and also names the section correctly.
%%%%%%%%%%%%%%%%%%%%%%%%%%%%%%%%%%%%%%%%%%%%%%%%%%%%%%%%%%%%%%%%%%%%%
\bibliography{bibliography2}

\end{document}